\documentclass[12pt]{article}
\usepackage{amssymb}
\usepackage{amsmath}
\usepackage{setspace}
\usepackage{url}
\usepackage[dvips,letterpaper,margin=2cm,includefoot]{geometry}

\usepackage{graphicx}
\usepackage{dcolumn}
\usepackage{bm}
\usepackage{subfig}
\usepackage[percent]{overpic}

\makeatletter
\newcommand\ackname{Acknowledgements}
\if@titlepage
  \newenvironment{acknowledgements}{%
      \titlepage
      \null\vfil
      \@beginparpenalty\@lowpenalty
      \begin{center}%
        \bfseries \ackname
        \@endparpenalty\@M
      \end{center}}%
     {\par\vfil\null\endtitlepage}
\else
  \newenvironment{acknowledgements}{%
      \if@twocolumn
        \section*{\abstractname}%
      \else
        \small
        \begin{center}%
          {\bfseries \ackname\vspace{-.5em}\vspace{\z@}}%
        \end{center}%
        \quotation
      \fi}
      {\if@twocolumn\else\endquotation\fi}
\fi
\makeatother

\begin{document}


\title{How Web 1.0 Fails: The Mismatch Between Hyperlinks and Clickstreams}

\author{Lingfei Wu\footnote{Department of Media and Communication, City University of Hong Kong.} \and Robert Ackland\footnote{Corresponding Author. Australian Demographic and Social Research Institute, The Australian National University.  Postal address: Coombs Building (no. 9), The Australian National University, ACT 0200, AUSTRALIA. TEL: +61 02 6125 0312; FAX: +61 02 6125 2992, \texttt{robert.ackland@anu.edu.au}}}

\date{\today}

\maketitle

\begin{abstract}
The core of the Web is a hyperlink navigation system collaboratively set up by webmasters to help users find desired information.  While it is well known that search engines are important for navigation, the extent to which search has led to a mismatch between hyperlinks and the pathways that users actually take has not been quantified.  By applying network science to publicly-available hyperlink and clickstream data for approximately 1000 of the top websites, we show that the mismatch between hyperlinks and clickstreams is indeed substantial.  We demonstrate that this mismatch has arisen because webmasters attempt to build a global virtual world without geographical or cultural boundaries, but users in fact prefer to navigate within more fragmented, language-based groups of websites. We call this type of behavior ``preferential navigation" and find that it is driven by ``local" search engines.

\end{abstract}





\newpage

\section{Introduction}

Invented by Tim Berners Lee in 1991, the World Wide Web is regarded as the ``largest human information construct in history'' (\url{http://webscience.org/webscience.html}).  The Web is commonly understood to have had three overlapping phases of development or eras. Under Web 1.0, webmasters provide content that is consumed by users, while Web 2.0 blurs the distinction between webmasters and users, with blogging tools, social network sites (e.g. Facebook) and micro-blog services (e.g. Twitter) enabling non-technical people to both produce and consume content (``prosumption'')\cite{ritzer2010production}. Web 3.0, or the Semantic Web, involves technologies that make the Web more machine-readable, leading to a ``web of data'', which is an evolution of the Web 1.0 ``web of documents'' \cite{shadbolt2006semantic}.

A common feature of all three phases is the use of technologies to help people find web content. With Web 1.0, and to a lesser extent Web 2.0, the core enabling technology is the hyperlink, which allows users to efficiently move around the Web, while Web 3.0 envisages automated agents finding content on behalf of users by drawing on users' browsing habits.

The present paper uses novel data and methods to investigate the extent to which web navigation is based on hyperlinks.  We construct clickstream and hyperlink networks comprised of the same 980 websites.  In the clickstream network, a directed weighted edge between websites $i$ and $j$ indicates the percentage of global Web users who visited website $i$ and then immediately visited website $j$. The clickstream network thus shows the pathways that people are taking as they navigate the Web. In the hyperlink network, a directed unweighted edge between websites $i$ and $j$ indicates that $i$ hyperlinks to $j$, and hence the hyperlink network shows the pathways that webmasters are creating for users.

Our analysis reveals a substantial mismatch between the hyperlink and clickstream networks, allowing us to conclude that in navigating the Web, users tend to create their own pathways rather than following hyperlinks laid out by webmasters.  This mismatch between hyperlinks and clickstreams reveals different preferences of webmasters and users: while webmasters work collaboratively to build a fully-connected online society, users in fact only navigate within the fragmental parts of the Web that they favor, a behavior which we term ``preferential navigation".

\section{Related work}

The importance of hyperlinks to the Web has led to a large amount of research, with applied physics research into properties of hyperlink networks (e.g. power laws) and models that might explain their emergence \cite{barabasi1999emergence}, social scientific studies on the sociological motivations behind hyperlink creation \cite{ackland2011online} and the political implications of power laws on the Web \cite{hindman2003googlearchy}, and computer science research into how hyperlink structures can be used to improve web search \cite{kleinberg1999authoritative,page1997pagerank}. 

However, as noted in recent research on e-learning systems, hyperlinks are too simple to represent the rich connections between documents that are created by users' various online activities \cite{zhuge2009communities,zhuge2011semantic}. While both hyperlinks and clickstreams can be used to show connections between documents, the former reflects the preferences of a relatively small number of webmasters whereas the latter is the collaborative product of potentially massive numbers of users. Furthermore, the use of search engines, bookmarks, default home pages and historical viewing records means there are many ways through which users can move between websites when there is no hyperlink connection \cite{Qiu2005webtraffic,Meiss2010Bookmarks}.

Although clickstreams have been analyzed since the early days of the Web \cite{catledge1995browsing}, it is only in the past decade that they have been studied from a network perspective.  Much of this work has been focused on intra-website clickstream networks in order to improve the design of sites, including social networks \cite{schneider2009understanding}, tagging systems \cite{cattuto2007semiotic}, news sharing systems \cite{wu2007novelty} and citation systems \cite{bollen2009clickstream}.  Research into recommendation systems and mobile computation is also making use of clickstream network analysis \cite{kim2005clickstream,yamakami2006regularity}, and professional software for clickstream network analysis \cite{brainerd2001case} has also been developed.

Researchers have also studied inter-website clickstream flows \cite{Qiu2005webtraffic,Meiss2010Bookmarks} and this is the focus of the present study.  However a key difference between the present and earlier studies of inter-website clickstreams is that we jointly analyze the hyperlink and clickstream networks, allowing us to quantify the extent of the mismatch between these networks and identify underlying causes.


\section{Data and methods}

We selected the top 1,000 websites according to Google's traffic statistics in November 2010 (\url{http://www.google.com/adplanner/static/top1000/}). We then used Alexa (\url{http://www.alexa.com}) to retrieve the daily \textit{traffic} to these websites (which is averaged over three months) and also the daily \textit{clickstreams} between them.  The sum of clickstreams to a given site will be less than or equal to the traffic to that site, since clickstreams only refer to visits from the set of 1,000 websites, while traffic is \textit{all} visits to the site.  According our calculation based on Google's statistics, these 1,000 websites account for more than $96\%$ of global Web traffic during the period of the data collection.  The clickstream network contains 12,008 directed and weighted edges, where an edge between websites $i$ and $j$ indicates the percentage of global Web users who visited $j$ immediately after visiting $i$.

We then used VOSON (\url{http://voson.anu.edu.au/}), which is software for hyperlink network construction and anlaysis created by one of the authors, to construct a hyperlink network where a directed and unweighted edge between websites $i$ and $j$ indicates that $i$ contains a hyperlink to $j$.  The hyperlink network contains 15,907 edges.

Twenty sites were dropped due to a lack of data and thus our analysis is for the remaining 980 sites. Further, as Alexa reports a maximum of ten largest inbound and outbound clickstreams for each website, and the version of the VOSON web crawler that was used only collected a maximum of 1,000 outbound hyperlinks for each site, the two constructed networks necessarily do not include all of the clickstreams and hyperlinks between these 980 sites.
It is important to note that our data do not allow us to know exactly how a person navigates from website $i$ to $j$: navigation may occur either through a hyperlink, a search engine, or the user typing the URL into the browser (or equivalently, following a bookmark).

\section{Results}

\begin{center}
  \begin{figure*}[!ht]
  \centering
      \subfloat[\label{subfig-1:a}]{%
      \begin{overpic}[scale=0.4]{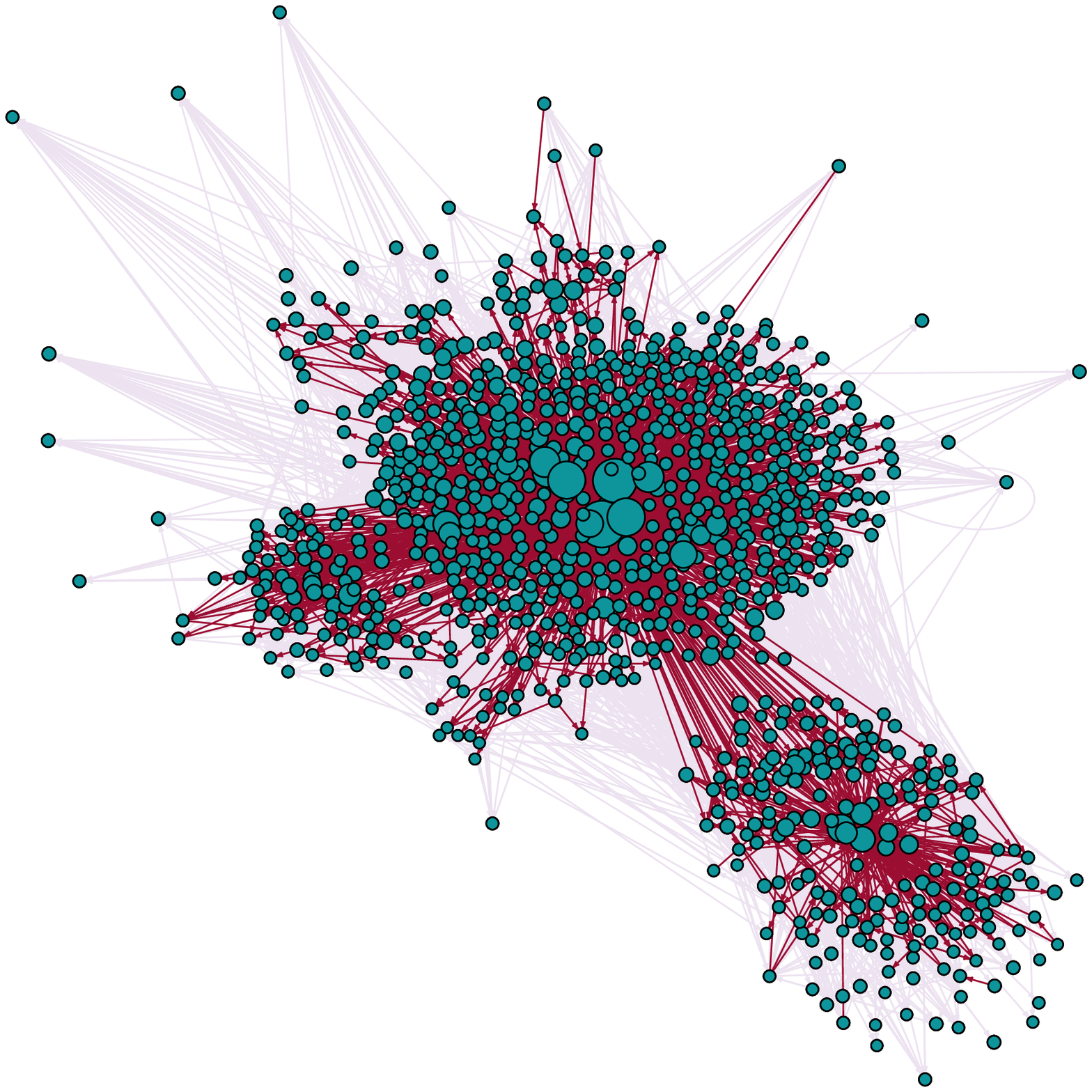}
      \end{overpic}
    }
       \subfloat[\label{subfig-2:b}]{%
      \begin{overpic}[scale=0.4]{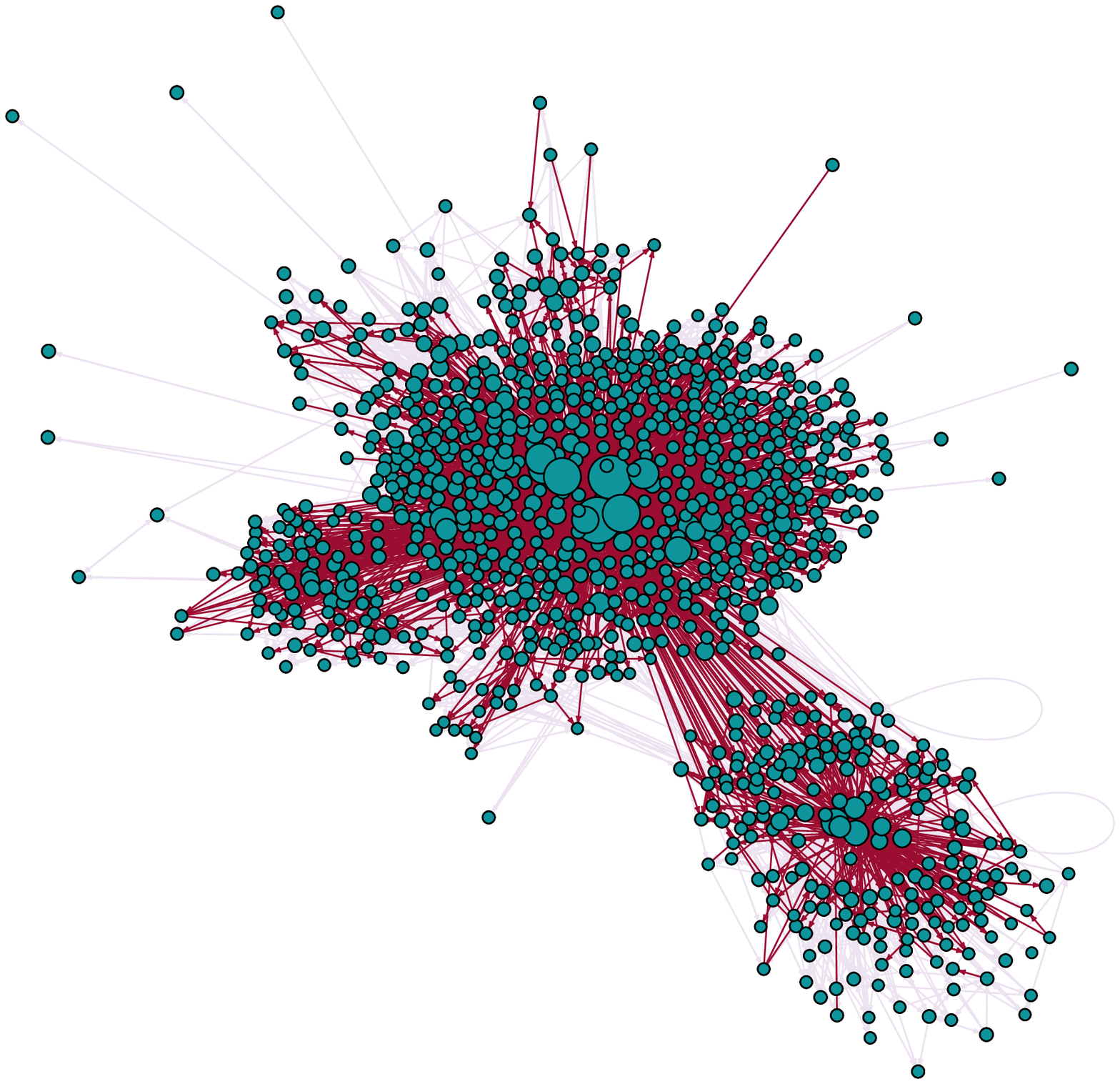}
      \end{overpic}
    }
\caption{The hyperlink network (a) and the clickstream network (b). The size of nodes denotes the logarithm of traffic. Edges that are common to both networks are drawn in red. The node layout method is Fruchterman-Reingold \cite{fruchterman1991graph}, and for ease of comparison, the two networks both have the node layout that was computed for the clickstream network.}
    \label{fig.1}
  \end{figure*}
 \end{center}

The hyperlink network and the clickstream network are shown in Figure \ref{fig.1}, with edges that are common to both networks drawn in red. We now explore differences between the two networks and the origins of the differences.


\subsection{\label{sec:1}The mismatch between hyperlinks and clickstreams}

The fact that the clickstream and hyperlink networks are comprised of the same nodes allows us to check for overlaps between the two sets of directed edges. Only 2,580 out of the 15,907 hyperlinks overlap with the 12,008 clickstreams, meaning that a large proportion of hyperlinks are ``useless" in the sense that they connect sites that did not exchange any traffic during the data collection period. The clickstreams transported by hyperlinks only account for $33\%$ of the sum of all clickstreams. This percentage gives an upper bound of the hyperlink-moderated clickstreams, since we collect more hyperlinks than clickstreams for each website. 

In other words, the actual proportion clickstreams driven by hyperlinks is likely to be even smaller. In Table.\ref{tab.1} we compare the hyperlink and clickstream networks in terms of several structural properties \cite{freeman1979centrality,Watts1998smallworld}. We note that the websites tend to cluster into small, local groups in the clickstream network (which has smaller global transitivity and density but is larger in average local transitivity). We conjecture that this structural difference reflects different navigation behaviors underlying the two networks; we explore this further in the following sections.

\begin{center}
\begin{table*}
\caption{Network statistics for the hyperlink and clickstream networks.}
\label{tab.1}
\begin{tabular}{lcc}
\hline
        & Hyperlink network & Clickstream network \\\hline
Number of nodes           &  980              &  980                \\
Number of edges           &  15907            &  12008              \\
Number of weak compnents  &  1                &  1                  \\
Global transitivity       &  0.117            &  0.051              \\
Average local transitivity&  0.410	          &  0.757              \\
Density                   &  0.017	          &  0.013              \\
Average path length       &  2.869            &  2.276              \\ \hline
\end{tabular}
\end{table*}
\end{center}

\subsection{\label{sec:2}The story behind the mismatch: Different preferences of webmasters and users}

To investigate the reason for the observed mismatch between the hyperlink and clicktream networks, we simulated user navigation in both networks using a label-propagation algorithm \cite{Raghavan2007linearAlgorithm}, which works as follows. Each website is initially assigned a unique label and then at every step of the simulation, each website adopts the most popular label in its neighbourhood. The process continues until there is no further change in labels, and websites with the same label are clustered into a community. 

\begin{figure*}[!ht]
    \centering
    \includegraphics[scale=0.5]{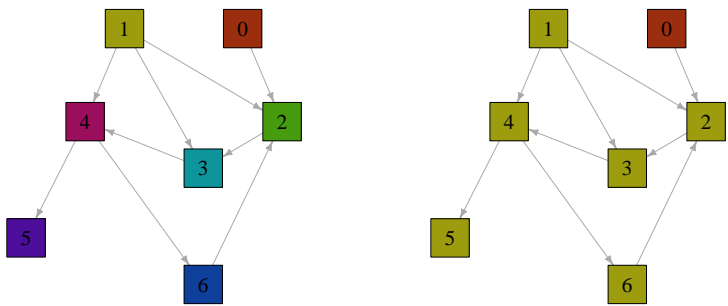}
    \caption{In the first step of the label-propogation simulation, each website is assigned a unique label (a). Then, each website adopts the most popular label in its neighbourhood. When the simulation stops, the websites with the same label are clustered into the same community (b).}
    \label{fig.2}
  \end{figure*}

To illustrate, website 2 in the example network shown in Figure \ref{fig.2} is visited by users from three websites, 0, 1, and 6. After several steps of simulation, website 6 is ``occupied" by users coming from website 1, therefore 2 will adopt the same label as 6 and 1. Eventually, websites 1, 6, 2 are clustered into a community, with website 0 belonging to another.  The label-propagation algorithm thus allows us to simulate of a group of users who share similar interests diffusing on the network and labeling all visited websites until coming to a website that has been ``occupied" by another group of users. The detected communities therefore correspond to the preferences of different users. 

\begin{figure*}[!ht]
    \centering
    \includegraphics[scale=0.5]{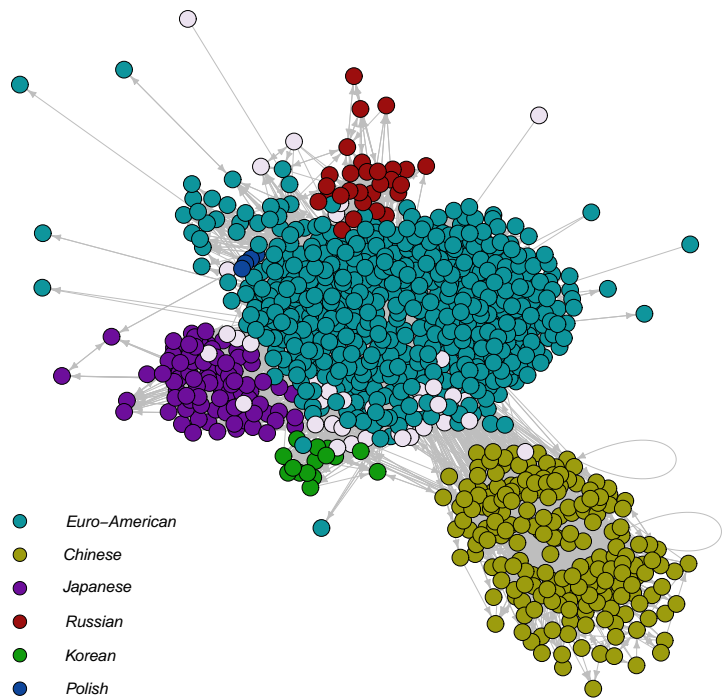}
    \caption{Six language-based communities detected from the clickstream network. The websites in different communities are shown in different colors. The nodes that are assigned to an incorrect community by the label propagation algorithm are plotted in white.}
    \label{fig.3}
  \end{figure*}

The simulation identified six communities from the clickstream network (Figure \ref{fig.3}), which broadly coincided with visual clusterings provided by the Fruchterman-Reingold algorithm. Websites in each community generally share the same language (with an accuracy rate of $96\%$, validated by human coding), and we therefore label the communities according to these languages: Polish community, Korean community, Russian community, Japanese community, Chinese community, and Euro-American Community (over 85\% of websites within this community are in English). As indicated by Table \ref{tab.2}, the Euro-American Community is the largest, accounting for 645 sites (66\% of the total) and 7,695 links (64\% of the total), while the smallest (Polish) consists of only 4 websites.

\begin{center}
\begin{table*}
\caption{Descriptive statistics of the clickstream network and its six communities. Clickstream is measured in number of unique visitors. ``APL" stands
for average path length.}
\label{tab.2}
\begin{tabular}{lccccc}
\hline
         Community & N of websites & N of edges & Density & APL  & Total daily clickstream \\\hline
Polish   Community &           4&     12&      1&      1&   $2.98 \times 10^{6}$\\
Korean   Community &          15&    114&  0.543&  1.457&   $2.29 \times 10^{6}$\\
Russian  Community &          28&    217&  0.287&  1.771&   $8.68 \times 10^{7}$\\
Japanese Community &          87&    899&  0.120&  1.884&   $2.34 \times 10^{8}$\\
Chinese  Community &         201&  2,058&  0.052&  1.994&   $7.96 \times 10^{8}$ \\
Euro-American Community&     645&  7,695&  0.019&  2.017&   $4.15 \times 10^{9}$ \\
Total                  &     980& 12,008&  0.013&  2.276&   $5.45 \times 10^{9}$ \\ \hline
\end{tabular}
\end{table*}
\end{center}

We conduct the simulation on the hyperlink network, but found that all websites were clustered into a single community.  The simulation therefore provides compelling evidence that webmasters and users exhibit very different preferences. The hyperlink network is constructed by webmasters, who link their websites to those they think visitors will also be interested in. If users followed the pathways set up by webmasters there would be a very high level of diffusion, as seen in the simulation conducted on the hyperlink network. However, the communities identified in the clickstream network suggest that the linking structure set up by the webmasters does not meet the requirements of users, who in fact prefer to navigate within local, language-based fragments of the Web. We call this behavior ``preferential navigation".

\subsection{\label{sec:3}Understanding web surfing behavior: Preferential navigation driven by local search engines}
In the previous section we showed that preferential navigation leads to the creation of language-based website communities. It was mentioned earlier that our clickstream data do not allow us to know the exact process by which a user visits two websites $i$ and $j$ in succession. There are three possibilities: (1) follow a hyperlink from $i$ to $j$; (2) enter the URL for $j$ directly into the browser after visiting $i$ (or equivalently, follow a bookmark for $j$); or (3) navigate from $i$ to a search engine, and then navigate to $j$ by clicking on a search result. We have already shown that (1) does not appear to account for a large proportion of clickstreams. The question we pose in the present section is: how important are search engines in user navigation, and what is their role in enabling preferential navigation?

To answer this question, we examined the role of search engines in the clickstream network. Firstly, we divided the clickstream network into sub-networks on the basis of language (this was done manually in light of the fact that the label propagation algorithm had a $4\%$ error rate). For each sub-network (the Polish community was excluded due to its small size) we calculated three centrality measures: degree, betweenness, and closeness.  According to \cite{freeman1979centrality}, these quantities reflect the importance of a node in different aspects: degree indicates the activity of a node, betweenness is a measure of a node's ability to control the flow in the network, and closeness shows a node's efficiency in resource transmission. We find that the five search engines, google.com (Euro-American), baidu.com (Chinese), yahoo.co.jp (Japanese), yandex.ru (Russian), and naver.com (Korean), have the highest values in terms of all three centrality measures. This finding clearly points to the predominant role of search engines in facilitating navigation.

The above analysis only reveals the static, structural importance of search engines in the clickstream network; to further investigate the role of search engines in navigation, we compared the clickstreams driven by these five search engines with the clickstreams moderated by hyperlinks. The five search engines moderate over $42\%$ of total clickstreams, with google.com accounting for over a half of the clickstreams moderated by these search engines, followed by baidu.com, yahoo.co.jp, yandex.ru, and naver.com. As mentioned above, hyperlinks only moderate $33\%$ of all the clickstreams. Thus it is reasonable to conclude that users rely more on search engines than hyperlinks in surfing the Web.

\begin{figure*}[!ht]
    \centering
    \includegraphics[scale=0.5]{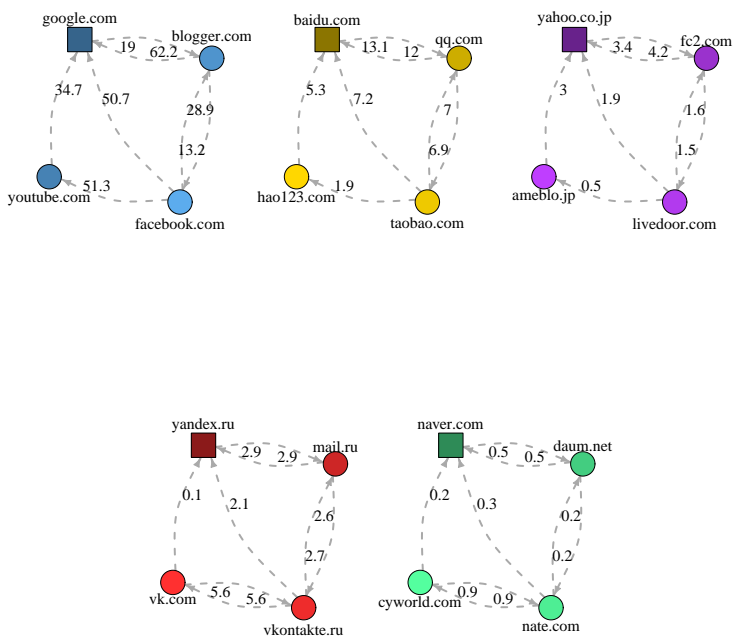}
    \caption{The ``popular pathways" in different communities. The squares shows the search engines as starting points and the circles denote other websites on the pathways. The weights on clickstreams indicate traffic measured in millions of unique users. The pathway colors correspond to the colors of communities in Figure \ref{fig.3}.}
    \label{fig.4}
  \end{figure*}

So how exactly do these search engines drive clickstreams? To address this question, we introduce a novel approach for analyzing clickstreams called ``popular-pathway-analysis". This approach is inspired by the ``maximizing chain strengths analysis" used in studies of food webs \cite{garlaschelli2003universal}. In each of the five communities, we start from a website ranking 1st in traffic according to Alexa's statistics. Then in each of the following steps, we choose the strongest (with the largest weight) outbound clickstream. We stop at the fourth step and draw the clickstream sub-network comprising the websites included in the selected clickstreams, which shows the ``most likely pathways" a typical user in the communities may take (Figure \ref{fig.4}).\footnote{Because we manually identified communities for our analyis of the importance of search engines in user navigation, this leads to an inconsistency between the clickstreams used for Figure \ref{fig.4} and those underlying Table \ref{tab.2} (where communities were automatically identified).  For example, the sum of the clickstreams for the four sites from the Korean community reported in Figure \ref{fig.4} is 3.7 million, which exceeds the total daily clickstreams for the entire community reported in Table \ref{tab.2} (2.29 million).  However, this inconsistency has no impact on our qualitative findings.} 

We found very similar circulations of clickstream in the five communities examined. In these circulations, users start from a ``local" search engine, and return to it repeatedly after visiting other websites. By using the term ``local" we mean that a search engine is only popular within a group of users speaking the same language. As search engines usually take into account the feedback (clicks) of users in ranking documents, a search engine frequently visited by users using the same language is more likely to recommend webpages in this language (e.g., baidu.com usually shows Chinese webpages on the first page), which in turn reinforces the preferences of users in choosing search engines.



\section{Discussion and Conclusion}

We present a novel approach for quantifying the mismatch between clickstreams and hyperlinks and apply this to data consisting of approximately 1000 of the top websites.  We find substantial evidence that users create their own pathways on the Web instead of following hyperlinks passively. We contend that this mismatch originates from the different preferences of webmaster and user: the former set up links to connect to each other's website collaboratively, leading to a highly connected hyperlink network, while the latter use local search engines to guide preferential navigation, which eventually results in a more fragmented, clustered network. It should be noted that although we focus on language in the current study, there are other factors (e.g., culture, commerce) that shape clickstream flows.

The findings in this study are relevant to several areas of web development. For example, hyperlink-based ranking algorithms may provide biased estimates of website relevance \cite{page1997pagerank,kleinberg1999authoritative} since they are derived from the hyperlink network structure created by webmasters, and our research has shown that web users do not tend to follow hyperlinks. Our findings on the extent of preferential navigation and the underlying causes are relevant to search engine companies who aim to become successful in more than one language-based community.

Our new approach can be used to benchmark and monitor the mismatch between clickstreams and hyperlinks as the Web evolves.  We predict that as the Web becomes increasingly intelligent, the hyperlink structure will gradually adapt to the clickstream structure, leading to a decrease of the mismatch. In Web 1.0, webmasters are the major constructors of the hyperlinks, and their limited information on user preferences leads to the mismatch. In the era of Web 2.0, users are able to hyperlink by themselves, for example linking from Facebook homepages to blogs. By encouraging users to contribute to the content of websites, webmasters incorporate user preference in setting up hyperlinks, and hence decrease the level of the mismatch \cite{Kim2004collaborativefiltering}. We can imagine that in the era of Web 3.0, as suggested by Tim Berners-Lee \cite{shadbolt2006semantic}, the mismatch between clickstreams and hyperlinks will continue to decrease, as the Web will be able to analyze users' historical surfing records and recommend appropriate websites, leading to the creation of hyperlinks that are more consistent with user web surfing preferences.

\begin{acknowledgements}
We thank Jonathan J. H. Zhu, Lexing Xie, Paul Thomas, and Hai Liang for providing comments on an earlier version of this paper.
\end{acknowledgements}

\bibliographystyle{unsrt}
\bibliography{test}

\section*{Authors}

Lingfei Wu is a doctoral student in the Department of Media and Communication, City University of Hong Kong. He can be contacted at \texttt{lingfeiwu2@student.cityu.edu.hk}.  Robert Ackland is an associate professor in the Australian Demographic and Social Research Institute, Australian National University.  He can be contacted at \texttt{robert.ackland@anu.edu.au}.

\end{document}